# Domination Problems in Nowhere-Dense Classes of Graphs

Anuj Dawar
University of Cambridge
Computer Lab, U.K.
anuj.dawar@cl.cam.ac.uk

Stephan Kreutzer
Oxford University
Computing Laboratory
kreutzer@comlab.ox.ac.uk

**Abstract**

We investigate the parameterized complexity of generalisations and variations of the dominating set problem on classes of graphs that are nowhere dense. In particular, we show that the distance-$d$ dominating-set problem, also known as the $(k,d)$-centres problem, is fixed-parameter tractable on any class that is nowhere dense and closed under induced subgraphs. This generalises known results about the dominating set problem on $H$-minor free classes, classes with locally excluded minors and classes of graphs of bounded expansion. A key feature of our proof is that it is based simply on the fact that these graph classes are uniformly quasi-wide, and does not rely on a structural decomposition. Our result also establishes that the distance-$d$ dominating-set problem is FPT on classes of bounded expansion, answering a question of Nešetřil and Ossona de Mendez.

## 1 Introduction

The dominating-set problem is among the most well-studied fundamental problems in algorithmic graph theory and complexity theory. Given a graph $G$ and an integer $k$, we are asked to determine whether $G$ contains a set $X$ of at most $k$ vertices such that every vertex of $G$ is either in $X$ or adjacent to a vertex in $X$. This is a classical NP-complete problem that has been intensively studied from the point of view of approximation algorithms and fixed-parameter tractability. A number of generalisations and variations of the dominating set problem have also been studied in this context. In particular, the *distance-d dominating-set problem* is one where we are given a graph $G$ and integer parameters $d$ and $k$ and we are to determine whether $G$ contains a set $X$ of at most $k$ vertices such that every vertex in $G$ has distance at most $d$ to a vertex in $X$. This problem, also known as the $(k,d)$-centre problem, has for instance been studied in [5] in connection with network centres and other clustering problems (see the references in [10]). It is clear that in the case $d=1$, this is just the dominating set problem. A number of other domination problems are considered in Section 5.

We are interested in investigating these problems from the point of view of fixed-parameter tractability. That is we are interested in algorithms for these problems that run in time $f(k) \cdot n^c$ (or $f(k+d) \cdot n^c$ in the case of the distance-$d$ dominating-set problem) where $n$ is the order of the graph $G$, $c$ is independent of the parameters $k$ and $d$ and $f$ is any computable function. Such algorithms are unlikely to exist in general, since the dominating-set problem is $W[2]$-complete (see [13, 15] for a general introduction to parameterised complexity, including definitions of FPT and $W[2]$). However, if we restrict the class of graphs under consideration, we can obtain efficient algorithms in the sense of fixed-parameter tractability, even though the problem remains NP-complete on the restricted class. We are interested in knowing how general we can make our graph classes while retaining fixed-parameter tractability. In this paper, we push the tractability frontier forward by showing that the distance-$d$ dominating-set problem as well as a number of other domination problems, are FPT on *nowhere-dense* classes of graphs. This generalises known results about the dominating set problem on $H$-minor free graphs, classes of graphs of *bounded expansion* and classes with *locally excluded minors*. Moreover, while the latter results relied heavily on graph structure theory, our proof depends on a simple combinatorial property of nowhere-dense classes and thus affords a great simplification to the proof. In the sequel, we will use the term *efficient algorithm* always to mean efficient in the sense of fixed-parameter tractability.

Classes on which efficient algorithms have previously been obtained for the dominating set problem include planar graphs where an algorithm with running time $\mathcal{O}(8^k n)$-time is given in [2] and graphs of genus $g$, where an $\mathcal{O}((4g+40)^k n^2)$-time algorithm is given in [14]. Improvements to the algorithms on planar graphs have subsequently been made, to $\mathcal{O}(4^{\sqrt[6]{34k}} n)$ in [1], to $\mathcal{O}(2^{27\sqrt{k}} n)$ in [17] and to $\mathcal{O}(2^{15.13\sqrt{k}} k + n^3 + k^4)$ in [16]. Efficient algorithms for distance-$d$ dominating sets are also known for planar graphs and map-graphs [10]. For the dominating set problem, efficient algorithms were shown for $H$-minor free graphs in [11]. The latter generalises the result for classes of graphs of bounded genus. More recently, Alon and Gutner gave a linear time parameterized algorithm for dominating sets on $d$-degenerate graphs running in time $k^{\mathcal{O}(dk)} n$ [3]. This is a further generalisation beyond $H$-minor-free classes. It should be noted that while all other classes mentioned above also admit an efficient algorithm for the distance-$d$ dominating-set problem, this is not the case for classes of degenerate graphs. Indeed, this problem is $W[2]$-hard, already on the class of 4-degenerate graphs.

Other generalisations of $H$-minor free classes that have been considered in the literature are classes with locally excluded minors [9] and classes of bounded expansion [19]. For the former, it follows from results of [9] that the distance-$d$ dominating-set problem is FPT. This is because the problem can be specified by a first-order formula (depending on $d$ and $k$), and it is proved in [9] that any property so specified admits is FPT on classes that locally exclude a minor. For classes of bounded expansion, Nešetřil and Ossona de Mendez [21] show that the dominating set problem is solvable in fixed-parameter linear time, while the question of whether the distance-$d$ dominating-set problem is FPT on such classes in one that they left open. Indeed, they point out that their method cannot be used to show that the distance-2 dominating-set problem is FPT on classes of bounded expansion. Our result settles this question as it implies the existence of an efficient algorithm for distance-$d$ dominating-set on bounded-expansion classes.

Our main results concern classes of graphs that are *nowhere dense*. This is a concept introduced by Nešetřil and Ossona de Mendez [18, 20] that generalises both locally excluded minors and bounded expansion in the sense that any class of graphs that either locally excludes a minor or has bounded expansion is also nowhere dense. Nešetřil and Ossona de Mendez show that nowhere-dense classes can be characterised by the property of being *uniformly quasi-wide* (see Section 2 for the defintions). The latter is a property introduced by Dawar [7, 8] in the study of preservation theorems in finite model theory. In the present paper we show that this property is by itself sufficient to establish that a class of graphs admits an efficient parameterized algorithm for distance-$d$ dominating set. The great advantage here is that this is a combinatorial property that is easy to state and yields a transparently simple algorithm. This should be contrasted with the algorithms [10, 11] on $H$-minor free graphs that heavily rely on graph structure theory.

We begin by establishing some basic terminology and notation in Section 2, and introduce nowhere-dense classes and uniformly quasi-wide classes of graphs. In Section 3 we examine the relationship between these two notions and extract the algorithmic content of the equivalence between them. This allows us, in Section 4, to exhibit an efficient parameterized algorithm for the distance-$d$ dominating set problem on nowhere dense classes. In Section 5, we explain how the same ideas can be carried over to a number of other parameterized problems that are defined in terms of domination in graphs.

## 2 Preliminaries

For a graph $G$ and vertices $u, v \in V(G)$, we write $\mathrm{dist}^G(u,v)$ for the distance (i.e. the length of the shortest path) from $u$ to $v$. We write $N_d^G(v)$ for the *d-neighbourhood* of $v$, i.e. the set of vertices $u$ in $V(G)$ with $\mathrm{dist}^G(u,v) \leq d$. Where the meaning is clear from the context, we may drop the superscript $G$. For positive integers $i < j$, we write $[i,j]$ for the set of integers $\{k : i \leq k \leq j\}$.

For a graph $G$ and a set of vertices $X \subseteq V(G)$, we write $G - X$ for $G[V(G) \setminus X]$, i.e. the subgraph of $G$ induced by the vertices $V(G) \setminus X$.

**2.1 Definition.** *Let $G$ be a graph and $d \in \mathbb{N}$.*

1. *A set $X \subseteq V(G)$ is $d$-scattered if for $u \neq v \in X$, $N_d(u) \cap N_d(v) = \varnothing$.*

2. *A set $X \subseteq V(G)$ $d$-dominates a set $W \subseteq V(G)$ if for all $u \in W$ there is a $v \in X$ such that $u \in N_d(v)$.*

3. *A set $X \subseteq V(G)$ is a $d$-dominating set if it $d$-dominates $V(G)$.*

We say that a graph $H$ is a *minor* of $G$ (written $H \preceq G$) if $H$ can be obtained from a subgraph of $G$ by contracting edges. An equivalent characterization (see [12]) states that $H$ is a minor of $G$ if there is a map that associates to each vertex $v$ of $H$ a non-empty *connected* subgraph $G_v$ of $G$ such that $G_u$ and $G_v$ are disjoint for $u \neq v$ and whenever there is an edge between $u$ and $v$ in $H$ there is an edge in $G$ between some node in $G_u$ and some node in $G_v$. The subgraphs $G_v$ are called *branch sets*.

We say that $H$ is a *minor at depth $r$* of $G$ (and write $H \preceq_r G$) if $H$ is a minor of $G$ and this is witnessed by a collection of branch sets $\{G_v \mid v \in V(H)\}$, each of which is contained in a neighbourhood of $G$ of radius $r$. That is, for each $v \in V(H)$, there is a $w \in V(G)$ such that $G_v \subseteq N_r^G(w)$.

The following definition is due to Nešetřil and Ossona de Mendez [18, 20].

**2.2 Definition** (nowhere dense classes). *A class of graphs $\mathcal{C}$ is said to be* nowhere dense *if for every $r$ there is a graph $H$ such that $H \npreceq_r G$ for all $G \in \mathcal{C}$.*

It is immediate from the definition that if $\mathcal{C}$ excludes a minor then it is nowhere dense. It is also not difficult to show that classes of bounded expansion and classes that locally exclude minors are also nowhere dense.

Nešetřil and Ossona de Mendez show an interesting relationship between nowhere dense classes and a property of classes of structures introduced by Dawar [7, 8] called *quasi-wideness*.

**2.3 Definition** (quasi-wide classes). *Let $s : \mathbb{N} \to \mathbb{N}$ be a function. A class $\mathcal{C}$ of graphs is* quasi-wide with margin $s$ *if for all $r \geq 0$ and $m \geq 0$ there exists an $N \geq 0$ such that if $G \in \mathcal{C}$ and $|G| > N$ then there is a set $S \subseteq V(G)$ with $|S| < s(r)$ such that $G - S$ contains an $r$-scattered set of size at least $m$.*
*We say that $\mathcal{C}$ is* quasi-wide *if there is some $s$ such that $\mathcal{C}$ is quasi-wide with margin $s$.*

We occasionally refer to a set $S$ as in the above definition as a *bottleneck* of $G$.

It turns out that if $\mathcal{C}$ is closed under taking induced subgraphs, then it is nowhere dense if, and only if, it is quasi-wide. For such classes, quasi-wideness is equivalent to the notion of uniform quasi-wideness defined below, which is the notion we will use in the present paper.

**2.4 Definition** (uniformly quasi-wide classes). *A class $\mathcal{C}$ of graphs is* uniformly quasi-wide *with margin $s$ if for all $r \geq 0$ and all $m \geq 0$ there exists an $N \geq 0$ such that if $G \in \mathcal{C}$ and $W \subseteq V(G)$ with $|W| > N$ then there is a set $S \subseteq V(G)$ with $|S| < s(r)$ such that $W$ contains an $r$-scattered set of size at least $m$ in $G - S$.*

We often write $s_\mathcal{C}$ for the margin of the class $\mathcal{C}$, and $N_\mathcal{C}(r, m)$ for the value of $N$ it guarantees for this class. We are only interested in classes $\mathcal{C}$ for which these functions are *computable*, and we tacitly make this assumption from now on.

We can now state the equivalence of the two central notions.

**2.5 Theorem** (Nešetřil-Ossona de Mendez). *Any class $\mathcal{C}$ of graphs that is closed under taking subgraphs, is quasi-wide if, and only if, it is nowhere dense.*

In Section 3, we will exhibit the algorithmic content of this equivalence by showing that in any nowhere-dense class, there is an efficient (in the sense of fixed-parameter tractability) algorithm that can find the bottleneck $S$ and the scattered set induced by its removal.

We end this section with some examples of quasi-wide classes.

**2.6 Example.**  1. Bounded-degree graphs. *The class of graphs $\mathcal{D}_d$ of valence at most $d$ is quasi-wide with margin $0$ and $N_{\mathcal{D}_d}(r, m) = (d-1)^r + d + 1$.*

2. $H$-minor free graphs. *The class of graphs excluding $H$ as a minor is quasi-wide with margin $k - 1$.*

# 3 Computing Bottlenecks and Scattered Sets

In this section, our aim is to extract the computational content of Theorem 2.5 stating the equivalence between nowhere dense classes and uniformly quasi-wide classes. In particular, we show that in any nowhere dense class $\mathcal{C}$, we can efficiently extract bottlenecks and scattered sets in any graph.

The first step is to show that in any uniformly quasi-wide class with margin $s$, we can compute, from $s(r)$ and $N_\mathcal{C}$, a bound on the order of the graphs that are excluded as minors of depth $r$.

**3.1 Lemma.** *If $\mathcal{C}$ is a uniformly quasi-wide class with margin $s$ and $h > N_\mathcal{C}(r+1, s(r+1)+1)$, then $K_h \npreceq_r G$ for any $G \in \mathcal{C}$.*

*Proof.* Suppose, for contradiction, that $K_h \preceq_r G$ and let $u_1, \ldots, u_h$ be such that the neighbourhoods $N_r^G(u_i)$ contain branch sets $G_1, \ldots, G_h$ witnessing this. Then, by the choice of $h$ and the definition of quasi-wideness, there is a set $S \subseteq V(G)$ with $|S| < s(r+1)$ such that $\{u_1, \ldots, u_h\}$ contains an $r+1$-scattered set $A$ of size $s(r+1)+1$ in $G-S$. Thus, since the branch sets are pairwise disjoint, there must be two distinct vertices $u_i, u_j \in A$ such that $S \cap G_i = S \cap G_j = \varnothing$. There is an edge between some vertex in $G_i$ and some vertex in $G_j$ (since they are branch sets witnessing $K_h \preceq_r G$). We thus have that $N_{r+1}(u_i) \cap N_{r+1}(u_j) \neq \varnothing$ even in $G - S$, contradicting the fact that $A$ is $r+1$-scattered. □

The other direction is based on the following theorem, stated in [8], though the proof is extracted from that of a weaker statement proved in [4].

**3.2 Theorem.** *[8] For any $h, r, m \geq 0$ there is an $N \geq 0$ such that if $G$ is a graph with more than $N$ vertices then*

1. *either $K_h \preceq_{r+1} G$; or*

2. *there is a set $S \subseteq V(G)$ with $|S| \leq h - 2$ such that $G - S$ contains an $r$-scattered set of size $m$.*

Indeed, the bound $N$ is computable as a function of $h$, $r$ and $m$. To be precise, let $R$ be the function guaranteed by Ramsey's theorem so that for any set $A$ with $|A| > R(x, y, z)$ any colouring of the $y$-tuples from $A$ with $x$ distinct colours yields a homogeneous subset of size at least $z$. Let $b_h(x) = R(k+1, h, (h-2)(x+1))$ and let $c_h(x) = R(2, 2, b_h^{h-2}(x))$ where $b_h^i(x)$ denotes the function $b_h$ iterated $i$ times. Then, it follows from the construction in [4] that taking $N(h, r, m) = c_h^r(m)$ (i.e. $c_h$ iterated $r$ times) suffices for the proof of Theorem 3.2.

It follows from the above that if $\mathcal{C}$ is a nowhere-dense class of graphs with a computable function $h$ such that $K_{h(r)} \npreceq_r G$ for any $G \in \mathcal{C}$, then $\mathcal{C}$ is quasi-wide with computable margin $s$ and a computable function $N_\mathcal{C}$. We now show that in this case, we can compute rather more. That is, given a graph $G \in \mathcal{C}$ and a set $W \subseteq V(G)$ with $|W| > N(h(r), r, m)$, we can find, in time $\mathcal{O}(|G|^2)$, a set $S$ and a subset $A \subseteq W$ of at least $m$ elements so that in $G - S$, $A$ is $r$-scattered. This is formalised in the lemma below, which relies on extracting the algorithmic content of the proofs in [4].

**3.3 Lemma.** *Let $\mathcal{C}$ be a nowhere-dense class of graphs and $h$ be the function such that $K_{h(r)} \npreceq_r G$ for all $G \in \mathcal{C}$. The following problem is solvable in time $\mathcal{O}(|G|^2)$.*

| Input: | $G \in \mathcal{C}$, $r, m \in \mathbb{N}$, $W \subseteq V(G)$ with $|W| > N(h(r), r, m)$ |
|---|---|
| Problem: | compute a set $S \subseteq V(G)$, $|S| \leq h(r) - 2$ and a set $A \subseteq W$ with $|A| \geq m$, such that in $G - S$, $A$ is $r$-scattered. |

*Proof.* In what follows, we write $h$ for $h(r)$ and $N$ for $N(h, r, m)$.

The proof proceeds by constructing sequences of sets of vertices $W_0 \supseteq W_1 \supseteq \cdots \supseteq W_r$ and $S_0 \subseteq S_1 \subseteq \cdots \subseteq S_r = S$ such that for all $i$,

1. $|S_i| < h - 1$

2. $W_i$ is $i$-scattered in $G - S_i$

3. $c_h^{r-i}(m) < |W_i| \leq N + 1$

4. for all $v \in S_i$ and $u \in W_i$ there is a $w \in N_i^{G-S_i}(v)$ such that $\{v, w\} \in E(G)$.

For $i = 0$, we take $S_0 = \varnothing$ and $W_0 = W$. It is clear that all four conditions are satisfied.

Suppose that $S_i$ and $W_i$ have been constructed. We define a graph $G'$ on the set of vertices $W_i$ by putting an edge between $u$ and $v$ if there is an edge in $E(G)$ between some vertex in $N_i^{G-S_i}(u)$ and $N_i^{G-S_i}(v)$. Since $K_h \not\preceq G$, $G'$ cannot contain a $h$-clique and thus as $|W_i| > c_h^{r-i}(m) = R(2, 2, b_h h - 2(c_h^{r-(i+1)}(m)))$, $G'$ contains an independent set $I$ with $|I| > b_h^{h-2}(c_h^{r-(i+1)}(m))$, which can be found by a greedy algorithm. Note that $G'$ can be constructed from $G$ in linear time, thus $I$ is found in quadratic time.

The proof of Lemma 5.2 in [4] then guarantees that as long as $K_h \not\preceq_{i+1} G$ we can find $W_{i+1} \subseteq I$ and $S_{i+1}$ satisfying the four conditions above. This is because the condition $K_h \not\preceq_{i+1} G$ guarantees that there is a (possibly empty) set of vertices $Z$ in $G - S_i$ with $|S_i \cup Z| < h - 1$ and a set $J \subseteq I$ with $|J| > c_h^{r-(i+1)}(m)$ such that $N_{i+1}^{G-S_i}(u) \cap N_{i+1}^{G-S_i}(v) = Z$ for each $u, v \in J$. Moreover, the choice of size bounds ensures that $Z$ can be found by a greedy algorithm. We start by taking $Z_0 := \varnothing$ and $I_0 := I$. Once $Z_j$ has been constructed (for $j < h - 2$), we check if there is a vertex $z$ such that there are more than $b_h^{h-2-j}(c_h^{r-(i+1)}(m))$ elements $v \in I_j$ such that $z \in N_{i+1}^{G-(S_i \cup Z_j)}(v)$. If there is, we take $I_{j+1}$ to be the set of such elements $v$ and $Z_{j+1} := Z_j \cup \{z\}$. This process is guaranteed to halt within at most $h - 2$ steps, at which point a greedy algorithm can find a set $J$ with at least $c_h^{r-(i+1)}(m)$ vertices with $N_{i+1}^{G-(S_i \cup Z_j)}(u) \cap N_{i+1}^{G-(S_i \cup Z_j)}(v) = \varnothing$, as otherwise we will have found $K_h$ as a minor of $G$ at depth $i + 1$. Thus, we take $S_{i+1} = S_i \cup Z$ and $W_{i+1} = J$ to satisfy the four conditions above. $\square$

The algorithm for distance-$d$ dominating set we present in Section 4 below makes repeated use of the procedure defined above to recursively reduce the problem of finding a distance-$d$-dominating set of size $k$ in a graph down to the size $N := N_{\mathcal{C}}(d, (k+2)(d+1)^s)$, at which point an exhaustive search is performed. The running time of the algorithm is thus exponential in $N$ (which only depends on the parameters), and cubic in $|G|$. On the other hand, the exact parameter dependence of the algorithm depends on the function $h$, which is determined by the class of structures $\mathcal{C}$. However, even for simple classes $\mathcal{C}$, where $h$ is linear, or constant, $N$ may be a rather fast-growing function of $k$ and $d$, as it is defined in terms of iterations of the Ramsey function $R$. On the other hand, as we saw in Example 2.6, there are quasi-wide classes, such as classes of graphs of bounded degree, where $N$ can be bounded by a simple exponential.

The property of being quasi-wide can be seen as stating the existence of weak separators. Recall that a set $S$ is a separator of a set of vertices $W$ in a graph $G$ if in the graph $G - S$, $W$ is split into non-empty disjoint parts with no path between them. It is known, for instance, that if $G$ is a graph of treewidth at most $h$, for any set $W$ of vertices, we can find a separator $S$ of $W$ with $|S| \leq h + 1$. Now, nowhere-dense graphs are a generalisation of classes of $H$-minor free graphs which include, in particular, classes of bounded treewidth. While we cannot hope for the separator property of the form that holds on bounded treewidth classes to hold in nowhere-dense classes, uniform quasi-wideness does show us that we can find a small set $S$ that splits $W$ into parts so that there are no *short* paths between them.

## 4   Distance-$d$-Dominating Set

In this section, we show that the distance-$d$-dominating set problem is fixed-parameter tractable on any nowhere-dense class $\mathcal{C}$ of graphs, with parameter $d + k$. Throughout the remainder of this section, fix a class $\mathcal{C}$ that is uniformly quasi-wide with margin $s_C$ and let $N_C(r, m)$ be as in Definition 2.4.

We consider a more general form of the problem. We are given a graph $G$ and a set $W \subseteq V(G)$ of vertices and we are asked to determine whether there is a set $X$ in $G$ of at most $k$ vertices that $d$-dominates $W$. We begin with the observation that this problem, when parameterized by $k$, $d$ and *the size of $W$* is FPT on the class of all graphs.

**4.1 Lemma.**  *The following problem is fixed parameter tractable.*

> *Input:*      A graph $G$, $W \subseteq V(G)$, $k, d \geq 0$
> *Parameter:*  $k + d + |W|$
> *Problem:*    Are there $x_1, \ldots, x_k \in V(G)$ such that $W \subseteq \bigcup_i N_d(x_i)$?

*Proof.* Consider any partition of $W$ into $k$ sets $W_1, \ldots, W_k$. For each $i \in [1, k]$, define the set $X_i := \bigcap_{w \in W_i} N_d(w)$. That is, $X_i$ is the set of vertices that individually $d$-dominate the set $W_i$. Now, if each $X_i$ is non-empty, then we can find the dominating set we are looking for by choosing $x_i$ to be any element of $X_i$. Conversely, any set $\{x_1, \ldots, x_k\}$ that $d$-dominates $W$ detemines a partition $W_1, \ldots, W_k$ such that $x_i \in X_i$.

The algorithm proceeds by considering each partition of $W$ into $k$ sets in turn (note that the number of such partitions is less than $k^{|W|}$). For each partition, we compute the sets $X_i$ by computing $N_d(w)$ for each $w \in W$ and taking appropriate intersections. This takes time $\mathcal{O}(d \cdot |W| \cdot |G|)$. The total running time is therefore $\mathcal{O}(d \cdot |W| \cdot k^{|W|} \cdot |G|)$ □

Now we want to consider the case where the size of $W$ is not part of the parameter, but $G$ is chosen from the nowhere-dense class $\mathcal{C}$. We show that in this case, we can find a set $W' \subseteq W$ whose size is bounded by a function of the parameters $k$ and $d$ such that $G$ contains a set of size $k$ that $d$-dominates $W$ if, and only if, it contains such a set that $d$-dominates $W'$. This will then allow us to use Lemma 4.1 to get the desired result.

For now, fix $k$ and $d$, and let $s := s_\mathcal{C}(d)$ and $N := N_\mathcal{C}(d, (k+2)(d+1)^s)$. That is, for any $G \in \mathcal{C}$ and $W \subseteq V(G)$ with $|W| > N$, we can find $S \subseteq V(G)$ and $A \subseteq W$ such that $|S| \leq s$, $|A| \geq (k+2)(d+1)^s$ and $A$ is $d$-scattered in $G - S$. We claim that, in this case, we can efficiently find an element $a \in A$ such that $G$ contains a set of size $k$ that $d$-dominates $W$ if, and only if, there is such a set that $d$-dominates $W \setminus \{a\}$. We formalise this statement in the lemma below.

**4.2 Lemma.** *There is an algorithm, running in time $f(k,d)|G|^2$ for some computable function $f$, that given $G \in \mathcal{C}$ and $W \subseteq V(G)$ with $|W| > N$ returns a vertex $w \in W$ such that for any set $X \subseteq V(G)$ with $|X| \leq k$, $X$ $d$-dominates $W$ if, and only if, $X$ $d$-dominates $W \setminus \{w\}$.*

*Proof.* By Lemma 3.3, we can find, in time $\mathcal{O}(|G|^2)$, $S \subseteq V(G)$ and $A \subseteq W$ such that $|S| \leq s$, $|A| \geq (k+2)(d+1)^s$ and $A$ is $d$-scattered in $G - S$. Let $S = \{t_1, \ldots, t_s\}$ and, for each $a \in A$, we compute the *distance vector* $\boldsymbol{v}_a = (v_1, \ldots, v_s)$ where $v_i = \text{dist}(a, t_i)$ if this distance is at most $d$ and $v_i = \infty$ otherwise. Note that there are, by construction, at most $(d+1)^s$ distinct distance vectors. Since $|A| \geq (k+2)(d+1)^s$, there are $k+2$ distinct elements $a_1, \ldots, a_{k+2} \in A$ which have the same distance vector. We claim that taking $w := a_1$ satisfies the lemma.

**Claim.** *For any set $X \subseteq V(G)$ with $|X| \leq k$, $X$ $d$-dominates $W$ if, and only if, $X$ $d$-dominates $W \setminus \{a_1\}$.*

The direction from left to right is obvious. Now, suppose $X$ $d$-dominates $W \setminus \{a_1\}$. Consider, the sets $A_i := N_d^{G-S}(a_i)$ for $i \in [2, k+2]$. These sets are, by construction, mutually disjoint. Since there are $k+1$ of these sets, at least one of them, say $A_j$, does not contain any element of $X$. However, since $a_j \in W \setminus \{a_1\}$ there is a path of length at most $d$ from some element $x$ in $X$ to $a_j$. This path must, therefore, go through an element of $S$. Since $\boldsymbol{v}_{a_1} = \boldsymbol{v}_{a_j}$, we conclude that there is also a path of length at most $d$ from $x$ to $a_1$ and therefor $X$ $d$-dominates $W$.

For the complexity bounds, note that all the distance vectors can be computed in time $\mathcal{O}(|S| \cdot |A| \cdot |G|)$. This is $f(k,d)|G|$ for a computable $f$. Adding this to the $\mathcal{O}(|G|^2)$ time to find $S$ and $A$ gives us the required bound. □

We now state the main result of this section.

**4.3 Theorem.** *The following problem is fixed-parameter tractable for any nowhere-dense class $\mathcal{C}$ of graphs.*

> DISTANCE-$d$-DOMINATING SET
> *Input:* A graph $G \in \mathcal{C}$, $W \subseteq V(G)$, $k, d \geq 0$
> *Parameter:* $k + d$
> *Problem:* Determine whether there is a set $X \subseteq V(G)$ of $k$ vertices which $d$-dominates $W$.

*Proof.* The algorithm proceeds as follows. Compute $s := s_\mathcal{C}(d)$ and $N := N_\mathcal{C}(d, (k+2)(d+1)^s)$. As long as $|W| > N$, use the procedure from Lemma 4.2 to choose an element $w \in W$ that may be removed. Once $|W| \leq N$, use the procedure from Lemma 4.1 to determine whether the required dominating set exists. □

This concludes the proof of Theorem 4.3. The following corollaries are immediate.

**4.4 Corollary.** *The dominating set problem is fixed-parameter tractable on any nowhere-dense class.*

This generalises the known results for the dominating set problem on classes of bounded expansion and classes that locally exclude a minor.

**4.5 Corollary.** *The distance-$d$ dominating set problem is fixed-parameter tractable with parameter $k + d$ on any class of graphs of bounded expansion, where $k$ is the size of the solution.*

This answers a question of Nešetřil and Ossona de Mendez who show that the dominating set problem is fixed-parameter tractable on such classes and ask whether the same could be true for the distance-2 dominating set problem.

## 5 Other Domination Problems

Among problems that are fixed-parameter intractable, dominating set and its variants play an important role. For instance, in the Compendium of Parameterized Problems [6], a number of problems are given which are known to be hard in general, but tractable on planar graphs. Virtually all of them are variations on the theme of finding dominating sets. In this section we show that all of these problems and, in many cases, their harder "distance-$d$" versions remain fixed-parameter tractable on nowhere-dense classes of graphs, which greatly generalises the results on planar graphs. We refer to [6] for formal definitions of the problems below and references to the literature where they were first studied.

The first type of problems we look at are dominating set problems with additional requirements for connectivity, such as CONNECTED DOMINATING SET where we are to compute a dominating set which induces a connected sub-graph. We study its generalisation to $d$-domination defined as follows.

CONNECTED DISTANCE-$d$-DOMINATING SET
*Input:* Graph $G$, $k, d \in \mathbb{N}$
*Parameter:* $k + d$
*Problem:* Is there a subset $X \subseteq V(G)$ with $|X| = k$ such that $X$ $d$-dominates $G$ and $G[X]$ is connected?

To see that this problem is FPT on nowhere-dense classes of graphs we first show that the following problem is fixed-parameter tractable on all graphs.

**5.1 Lemma.** *The following problem is fixed-parameter tractable.*

> *Input:* A graph $G \in \mathcal{C}$, $X_1, \ldots, X_k \subseteq V(G)$ pairwise disjoint, $k \geq 0$
> *Parameter:* $k$
> *Problem:* Are there vertices $x_1 \in X_1, \ldots, x_k \in X_k$ such that $G[x_1, \ldots, x_k]$ is connected?

*Proof.* Suppose $x_1, \ldots, x_k$ is a solution to the problem, i.e. $x_i \in X_i$ for all $i$ and $G[x_1, \ldots, x_k]$ is connected. Let $T'$ be a spanning tree of $G[x_1, \ldots, x_k]$. Hence, $T'$ is a witness showing that there exists an abstract tree $T$ on $k$ vertices $v_1, \ldots, v_k$ so that for each $v_i$, $1 \leq i \leq k$, there is an $x_i \in X_i$ and if $\{v_i, v_j\} \in E(T)$ then $\{x_i, x_j\} \in E(G)$. Conversely, if we can find such a tree $T$ then it induces a solution to the problem.

This suggests the following algorithm. For each tree $T$ on vertices $v_1, \ldots, v_k$: for $1 \leq i \leq k$ let $N_i := \{j : \{v_i, v_j\} \in E(T)\}$ be the indices of neighbours of $v_i$ in $T$. We can test if there are vertices $x_1 \in X_1, \ldots, x_k \in X_k$ such that for all $1 \leq i \leq k$ and all $j \in N_i$ there is an edge between $x_i$ and $x_j$ as follows. For $1 \leq i \leq k$ let $Y_i \subseteq X_i$ be the set of vertices $x \in X_i$ such that $x$ has a neighbour in $X_j$ for all $j \in N_i$. Now, while there is an $1 \leq i \leq k$, a $j \in N_i$ and a vertex $v \in Y_i$ without neighbour in $Y_j$, we remove $v$ from $Y_i$. Clearly, this iteration reaches a fixed point after at most $k^2 \cdot n$ iterations. Once the fixed point has been reached we simply check if all $Y_1, \ldots, Y_k$ are non-empty and if so, as the $X_i$ are pairwise disjoint and, hence, so are the $Y_i$, we can easily find a set $x_1 \in Y_1, \ldots, x_k \in Y_k$ such that $\{x_i, x_j\} \in V(G)$ for all $1 \leq i \leq k$ and $j \in N_i$. □

Lemma 5.1 immediately implies the following modified version of Lemma 4.1: we first compute the sets $X_1, \ldots, X_k$ as in the proof of Lemma 4.1. Now, for each partition $\mathcal{B} := \{B_1, \ldots, B_l\}$ of $\{1, \ldots, k\}$ we compute $Y_1, \ldots, Y_l$ where $Y_i := \bigcap \{N_d(X_j) : j \in B_i\} \setminus \bigcup \{N_d(X_j) : j \notin B_i\}$. Clearly, these sets are pairwise disjoint and we apply the procedure from Lemma 5.1 to each of those. If we find vertices $y_1 \in Y_1, \ldots, y_l \in Y_l$ inducing a connected sub-graph, then this is a connected $d$-dominating set. Conversely, any connected dominating set induces such a partition.

**5.2 Lemma.** *The following problem is fixed parameter tractable.*

| | |
|---:|:---|
| Input: | A graph $G$, $W \subseteq V(G)$, $k, d \geq 0$ |
| Parameter: | $k + d + |W|$ |
| Problem: | Are there $x_1, \ldots, x_k \in V(G)$ such that $W \subseteq \bigcup_i N_d(x_i)$ and $G[x_1, \ldots, x_k]$ is connected? |

Now, in the proof of Theorem 4.3, if instead of Lemma 4.1 we use Lemma 5.2 we can show that CONNECTED DISTANCE-$d$-DOMINATING SET is FPT on nowhere-dense classes of graphs. Taking $d = 1$, this implies the result for CONNECTED DOMINATING SET. Similar methods can be used to show that the problem $d$-CONNECTED DISTANCE-$d$-DOMINATING SET is FPT on nowhere-dense classes. This is the problem of deciding if there is a $d$-dominating set $X$ of $k$ vertices which is $d$-connected. A set is said to be $d$-connected in a graph $G$ if it induces a connected subgraph in the graph $G^d$ obtained from $G$ by putting an edge between any two vertices that have distance at most $d$ in $G$. Finally, the same method shows that EFFICIENT DOMINATING SET is in FPT on nowhere-dense graph classes.

EFFICIENT DOMINATING SET

| | |
|---:|:---|
| Input: | Graph $G$, $k \in \mathbb{N}$ |
| Parameter: | $k$ |
| Problem: | Is there a subset $X \subseteq V(G)$ with $|X| = k$ such that $X$ is a dominating set and, in addition, all pairs $x \neq y \in X$ have distance at least 3? |

Further variations of domination problems studied in the literature are ANNOTATED DOMINATING SET and RED-BLUE DOMINATING SET. Annotated

dominating means that we are given a graph $G$ and $W \subseteq V(G)$ and want to find a set dominating $W$. The distance-$d$-version of this problem is what is solved by Theorem 4.3. Red-Blue Domination means that we are given a graph $G$ whose vertex set is partitioned into blue and red vertices and we want to dominate the blue vertices using red vertices only. Again its distance-$d$ version can be solved by the methods presented in Section 4.

Finally, we look at problems such as ROMAN DOMINATION, MAXIMUM MINIMAL DOMINATING SET, PERFECT CODE and DIGRAPH KERNEL. If we are not interested in their distance-$d$-version than an even simpler algorithm than the one presented above can be used to show that these problems are in FPT on nowhere-dense classes of graphs, which we demonstrate using the ROMAN DOMINATION problem.

ROMAN DOMINATION
  *Input:* Graph $G$, $k \in \mathbb{N}$
  *Parameter:* $k$
  *Problem:* Is there a Roman domination function $R$ such that $\sum_{v \in V(G)} R(v) \leq k$?

A Roman domination of $G$ is a function $R : V(G) \to \{0, 1, 2\}$ such that for all $v \in V(G)$ if $R(v) = 0$ then there exists an $x \in N(v)$ such that $R(x) = 2$. To solve the problem on nowhere-dense classes of graphs we first compute a set $S \subseteq G$ such that $G \setminus S$ contains a 2-scattered set $A$ of size $2k + 1$. Clearly, for at least $k + 1$ vertices $v \in A$ we must have $R(v) = 0$ and hence one of their neighbours must be labelled by 2. However, this implies that at least one vertex in $S$ must be labelled by 2. As $|S|$ only depends on the parameter we can use this to define a recursive procedure whose depth and width only depend on the parameter.

The following theorem sums up what we have established so far. It is easily seen that INDEPENDENT SET and INDEPENDENT DOMINATING SET are FPT on nowhere-dense classes and our procedures presented before readily solve the problems. We refer to [6] for precise definitions of the problems.

**5.3 Theorem.** *The following problems are fixed-parameter tractable on nowhere-dense classes of graphs:* CONNECTED DOMINATING SET, CONNECTED $d$-DOMINATING SET, $d$-CONNECTED $d$-DOMINATING SET, ANNOTATED DOMINATING SET, EFFICIENT DOMINATING SET, MAXIMUM MINIMAL DOMINATING SET, ROMAN DOMINATION, RED-BLUE DOMINATING SET, INDEPENDENT SET, INDEPENDENT DOMINATING SET, PERFECT CODE, *and* DIGRAPH KERNEL.

These examples show that the distance-$d$-dominating set problem that we showed is tractable on nowhere-dense graph classes is actually representative of a whole class of similar problems which become tractable in this case. Several of these are known to be intractable on graph classes of bounded degeneracy.

To give an example of a problem which remains hard on nowhere-dense classes, consider the DIRECTED DISJOINT PATHS problem.

DIRECTED DISJOINT PATHS
  *Input:* Directed graph $G$, pairs $(s_1, t_1), \ldots, (s_k, t_k) \in V(G)^2$
  *Parameter:* $k$
  *Problem:* Does $G$ contain $k$ vertex disjoint paths $P_1, \ldots, P_k$ such that $P_i$ links $s_i$ and $t_i$?

This is known to be $W[1]$-hard even on acyclic digraphs and it is easy to see that DIRECTED DISJOINT PATHS can be reduced to the directed disjoint paths problem on graphs of degree at most 4 as follows. Let $G$ be a digraph and let $v \in V(G)$ be a vertex with in-neighbours $u_1, \ldots, u_l$ where $l > 1$. Let $T$ be a directed rooted tree of degree at most 3 which has $l$ leaves and where all edges are oriented towards the root. Now eliminate all incoming edges to $v$ and add the tree $T$ to $G$ identifying $v$ with the root of $T$ and $u_1, \ldots, u_l$ with the leaves of $T$. A similar procedure is used to eliminate outgoing edges of $v$. Applying this to all vertices in $G$ yields a graph $G'$ of degree at most 4 but which has $k$ disjoint paths between the pairs $(s_1, t_1), \ldots, (s_k, t_k)$ if, and only if, such paths exist in $G$. Since the class of graphs of degree at most 4 is nowhere dense, this shows the problem is hard on such classes as in the general case.

# 6   Conclusion and Further Work

The aim of the paper is to initiate an algorithmic study of graph classes which are nowhere dense. The examples above, including the dominating set problem and the more general distance-$d$ dominating set, or $(k, d)$-centre, problem, demonstrate that a certain class of important algorithmic problems become fixed-parameter tractable on classes which are nowhere dense. One of the main advantages of these results over known algorithms for these problems on classes excluding a fixed minor is that our algorithms are elementary and do not rely on deep results and methods from graph minor theory.

An obvious direction for further research is to investigate what other problems might become tractable on nowhere-dense classes of graphs. Also, it would be interesting to compare nowhere-dense classes of graphs to graph classes of bounded degeneracy. The two concepts are incomparable but both generalise classes excluding a fixed minor and we have already seen that there are examples of problems that become fixed-parameter tractable on nowhere-dense classes of graphs which are intractable on classes of graphs of bounded degeneracy.

Finally, it would be interesting to explore the extent of the algorithmic theory of nowhere-dense classes of graphs in terms of algorithmic meta-theorems. In particular, it would be very interesting if model-checking for first-order logic was FPT on nowhere-dense classes of graphs. This would establish a rich algorithmic theory of such classes. However, establishing such a result would require novel methods as we do not have a decomposition theory for nowhere-dense classes along the lines of what is used to establish the tractability of first-order logic in classes with locally excluded minors. We leave this for future research.